\documentclass[10pt]{article}
\usepackage{graphicx}
\usepackage{epsf}
\usepackage{psfig}
\usepackage{lscape}

\begin{document}
\title{Potential programs for high sensitivity FIR spectroscopy with SPICA}   
\author{Luigi Spinoglio, Anna Maria Di Giorgio, Paolo Saraceno\\
\it IFSI - INAF, Via Fosso del Cavaliere 100, 00133 Roma, Italy}    
\maketitle

\begin{abstract} 
We discuss the potential of high sensitivity mid-IR and far-IR spectroscopy to proof the physical properties of active nuclei and starburst regions of local and distant galaxies. For local galaxies, it will be possible to map the discs and ISM through the low ionization ionic lines and a variety of molecular tracers, such as OH, H$_{2}$O and high-J CO. At increasing distance, most of the ionic nebular lines (typical of stars and AGNs) are shifted into the FIR, making possible to compare the observed spectra with those predicted by different evolutionary scenarios. At the very high redshift of 10-15, sensitive mid-to-far-IR spectrometers, such as those planned to be at the focal plane  of the future SPICA mision, could be adequate to detect the H recombination lines excited in the HII regions around population III stars, if these stars happened to reside in large clusters of more than 10$^{5}$ members. 

\end{abstract}

\section{Introduction}

The spectroscopic observations of the Infrared Space Observatory (ISO) opened a new window for the study of the physical and chemical properties of IR-bright, ultraluminous infrared galaxies (ULIRGs) and active galactic nuclei (AGN). Most of our present knowledge of the mid-to-far infrared spectra of active and starburst galaxies is derived from ISO spectroscopy (Fischer et al 1999, Sturm et al. 2002, Verma et al. 2003). The mid-IR spectral range includes most of the fine-structure lines excited by the hard radiation produced by black hole accretion as well as those mainly excited by stellar ionization (e.g. Spinoglio \& Malkan 1992) and thus represents an essential tool to distinguish between the two processes, especially in obscured nuclei suffering severe dust extinction. In a complementary way, the far-IR range contains a large variety of molecular (OH, H$_{2}$O, high-J CO) and low excitation ionic/atomic transitions, in emission or in absorption, that can reveal the geometry and morphology of the circumnuclear and nuclear regions in galaxies. In particular these latter could trace the expected conditions of X-UV illuminated dusty tori (see e.g. Krolik \& Lepp 1989 for predictions of high-J CO lines), whose presence in type 2 active galaxies is foreseen to reconcile the type1/type2 dichotomy.

The far-IR spectra of local IR-bright and ULIRGs galaxies, as measured by ISO-LWS (Fischer et al.1999), showed an unexpected sequence of features, as can be seen in Fig.1, 
from strong [OIII]52, 88$\mu$m and [NIII]57$\mu$m line emission to detection of only faint [CII]157$\mu$m line emission and [OI]63$\mu$m in absorption. The [CII]157$\mu$m line in 15 ULIRGs (L$_{IR}\geq10^{12}L_{\odot}$) revealed an order of magnitude deficit compared to normal and starburst galaxies relative to the FIR continuum. Non-PDR components, such as dust-bounded photoionization regions, generating much of the FIR continuum but not contributing significant [CII] emission, can explain the [CII] deficiency. Such environments may also explain the suppression of FIR fine-structure emission from ionized gas and PAHs, and the warmer FIR colors of ULIRGs (Luhman et al 2003). 

Examples of far-infrared molecular lines detected by ISO include: OH at 34.6$\mu$m, 53.3$\mu$m, 79.1$\mu$m and 119$\mu$m connecting to the ground state, giving rise to absorption in most cases;
OH at  98.7$\mu$m, 163$\mu$m at higher energies, in emission;
H$_2$O at  73.5$\mu$m, 90$\mu$m, 101$\mu$m, 107$\mu$m, 180$\mu$m in emission or absorption.
From the absorption lines the gas column and the abundance can be derived, from the emission lines we can measure the gas cooling and constrain the temperature and density of the warm (50$\leq$T$\leq$500K) molecular gas.

LWS observations of Arp 220 show absorption in molecular lines of OH, H$_2$O, CH, NH, and NH$_3$, as well as in the [OI]63$\mu$m line and faint emission in the [CII]158$\mu$m line. The molecular absorption in the nuclear region is characterized by high excitation due to high infrared radiation density (Gonz\'alez-Alfonso et al 2004). The LWS spectrum of the prototype Seyfert 2 galaxy NGC1068, beside the expected ionic fine structure emission lines, shows the 79, 119 and 163$\mu$m OH rotational lines in emission, not in absorption as in every other galaxy yet observed. Modeling the OH lines suggests the emitting gas lies in small (0.1pc) and dense clouds ($\sim 10^4 {\rm cm^{-3}}$) most probably in the nuclear region and is potentially a signature of the torus (Spinoglio et al 2005). 

From the pioneering work of ISO it appears that both the mid-infrared and the far-infrared are powerful tools to study the physical conditions in galaxies, often obscured by dust. The future far-infrared spectrometer ESI (Swinyard 2006) at the focal plane of the helium cooled 3.5m class telescope of the SPICA mission (Nakagawa 2004, Onaka \& Nakagawa 2005), due to its high sensitivity, will be able to detect FIR lines in local galaxies, near-to-mid infrared lines and dust/PAH features in distant galaxies and possibly reveal the presence of large clusters of population III stars through H recombination lines.

\begin{figure}[ht]
\centerline{\includegraphics[width=10cm]{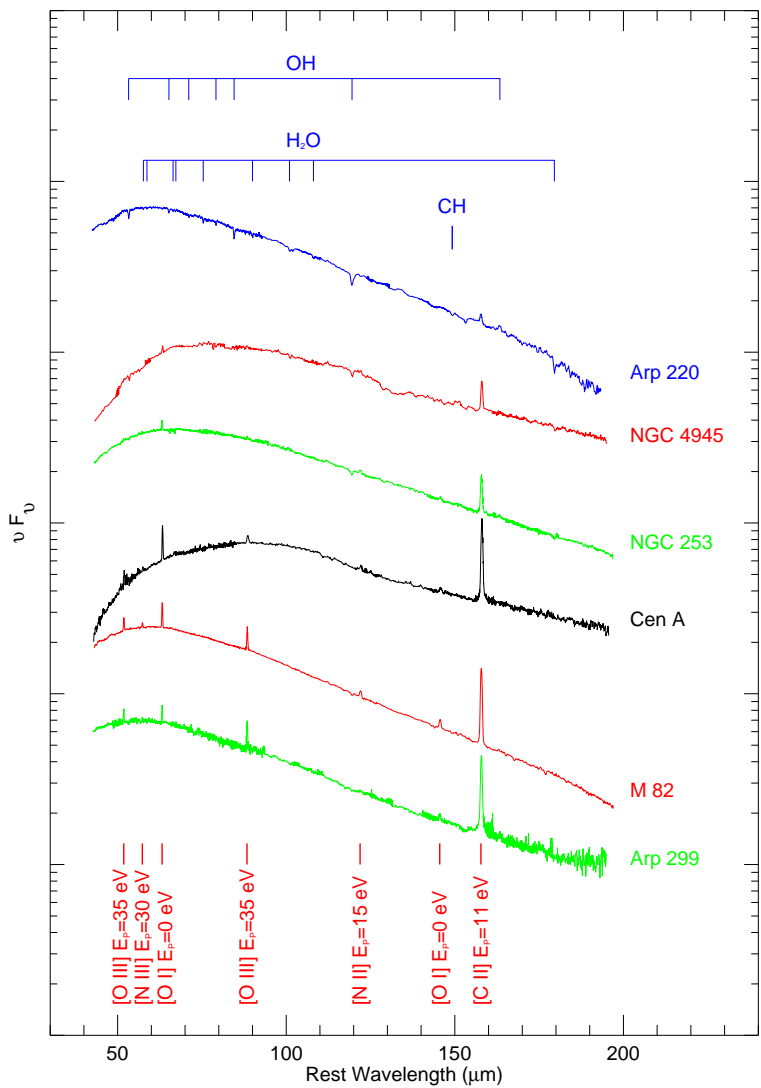}}
\caption{\footnotesize The ISO-LWS spectra of six IR-bright galaxies (adapted from Fischer
et~al 1999). The spectra are shifted and ordered vertically according to apparent excitation (not in order of relative luminosity or brightness).}\label{fig1}
\end{figure}

\section{Infrared line diagnostics through fine-structure emission lines}

Mid-IR and far-IR spectroscopy of fine-structure emission lines are powerful tools to understand the physical conditions in galaxies from the local universe to distant cosmological objects. 
Figure 2 shows the critical density (i.e. the density for which the rates of collisional and radiative de-excitation are equal) of each line as a function of the ionization potential of its ionic species, in three different redshift ranges, one for each frame, for which the rest-frame wavelength of the line is shifted in the far-infrared range. This diagram shows how these lines can measure two fundamental physical quantities (density and ionization) of the gas. 
In the figure are shown with different symbols the lines originated in different astrophysical emission regions at work in galaxies: the photodissociation regions (PDR), the stellar ionization in HII/starburst regions, the AGN ionization and the high ionization of coronal type emission. 
Choosing two lines along the abscissa and two along the ordinate, their ratios give, respectively, a good estimate of the ionization and the density of the gas in the region (see, e.g., Spinoglio \& Malkan 1992). 

\begin{figure}[ht]
\centerline{\includegraphics[angle=0,width=12cm]{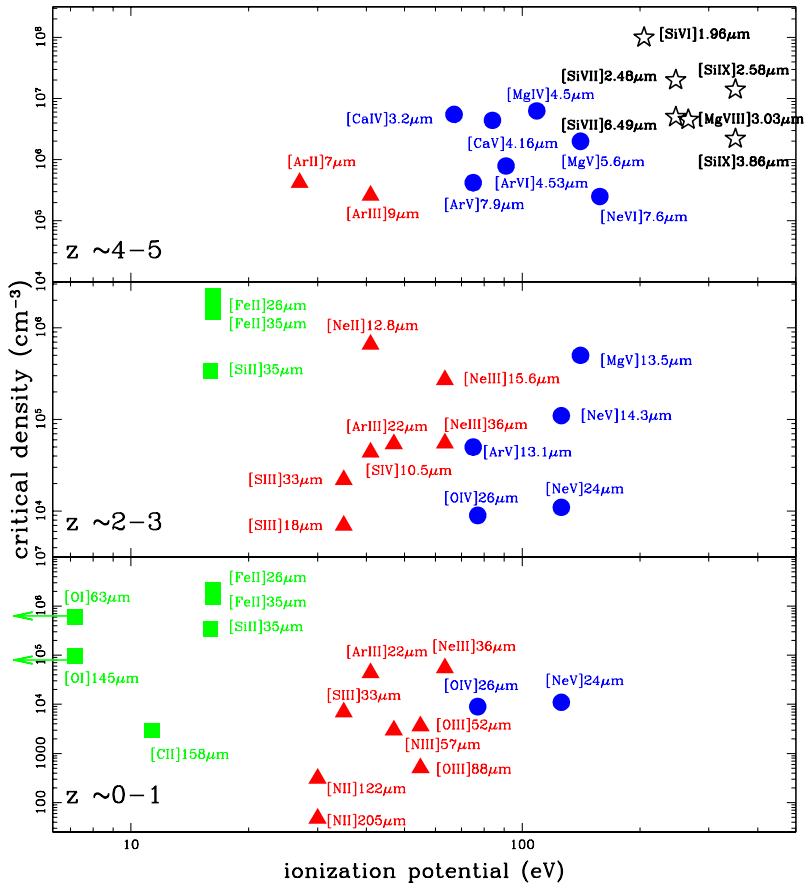}}
\caption{ \footnotesize Fine-structure lines occurring in the 30-200$\mu$m range at three different redshift ranges (the rest wavelength is reported near each line label). The lines are plotted  as a function of the ionization potential and their critical density. Different symbols are used for the lines from the photodissociation regions (PDR), shown as squares, the stellar/HII region lines, shown as triangles, the AGN lines, depicted with circles and the coronal lines, indicated with stars. The two [OI] lines have been plotted - for graphical reasons - at a ionization potential higher than their zero value.} \label{fig2}
\end{figure}

It appears from the figure that in the far-IR range, spectroscopy of the fine structure lines can probe in the Local Universe the two regimes of photodissociation (PDR) and stellar ionization. 
The stellar emission (e.g. in starburst galaxies) can be probed up to high z, using many lines in the rest-frame spectral range of 3 $\leq$ $\lambda(\mu m)$ $\leq$ 30.
The ionization from AGN can be probed in the redshift range 1$\leq$z$\leq$5 and
the coronal emission regions are probed by near-IR lines shifted into the far-IR at a redshift of z $\sim$ 5.

\section{Fine-structure line detection in Seyfert galaxies at increasing redshift}

The ISO spectrometers detected in the bright local Seyfert type 2 galaxy, NGC1068, many of the lines plotted in Fig.2 at flux levels of 5-200 $\times$ 10$^{-16}$ W m$^{-2}$ (Alexander et al 2000, Spinoglio et al 2005). This object contains both an active nucleus and a starburst.
Considering this galaxy as a template object, we computed the line intensities expected at redshifts ranging from 0.1 to 5.
For simplicity, we adopted an Einstein-De Sitter model Universe, with $\Omega_{\Lambda}$ = $\Omega_{vac}=0$ and  $\Omega_{M}$= 1, H$_{0}$=75 km s$^{-1}$ Mpc$^{-1}$. The luminosity distances have been derived using:
\begin{equation}
d_{L} (z)= 2c/H_{0} [1+z - \sqrt(1+z)]
\end{equation}

The results are reported in Table 1, where the line intensities are given in W m$^{-2}$. We have assumed that the line luminosities scale as the bolometric luminosity and we have chosen two cases: \\
A) a luminosity evolution proportional to the (z+1)$^{2}$, consistent with the {\it Spitzer} results at least up to redshift z=2 (Perez-Gonzalez et al 2005) (A values in the table);\\
B) no luminosity evolution (B values in the table).

Because the star formation process in galaxies was much more enhanced at z=1-2 than today, 
we consider reasonable to adopt the model with strong evolution (model A) for the stellar/HII region lines and the "no evolution" one (model B) for the AGN lines.

We note that the dependence on different cosmological models is not very strong. The popular model with $\Omega_{M}$= 0.27, $\Omega_{vac}$=0.73, H$_{0}$=71 km s$^{-1}$ Mpc$^{-1}$ shows greater dilutions, increasing with z, by factors of 1.5 for z=0.5 to 2.5 for z=5.

\begin{table*}[ht]
  \caption{Line intensity predictions in W/m$^{2}$ for NGC1068 at different redshifts}
  \footnotesize
\begin{center}
\begin{tabular}{@{}lccccccccccc@{}}
\hline
&&&&&&&&&&&\\
&  &[ArII] &[NeVI] &[ArIII] &[SIV] &[NeII] &[NeV] &[NeIII] &[SIII] &[NeV]&[OIV] \\
 $\lambda(\mu$m)      &   & 7.0 &7.6 &9.0 &10.5 &12.8 &14.3 &15.5 &18.7 &24.3 &25.9 \\

& Obs. &1.3e-15 & 1.1e-14 & 2.3e-15 &5.8e-15 &7.e-15 &9.7e-15 &1.6e-14 &4.e-15 &7.0e-15&1.9e-14 \\
\hline
z=0.1& A&- &-&-&-&-&-&-&-&-&2.4e-17\\
          &   B&- &-&-&-&-&-&-&-&-&2.0e-17\\
z=0.2  &A&-&-&-&-&-&-&-&-&2.5e-18&6.8e-18\\
        &     B&-&-&-&-&-&-&-&-&1.7e-18&4.7e-18\\
z=0.3  &A&-&-&-&-&-&-&-&-&1.3e-18&3.4e-18\\
        &     B&-&-&-&-&-&-&-&-&7.7e-19&2.0e-18\\
z=0.4  &A&-&-&-&-&-&-&-&-&8.0e-19&2.2e-18\\
        &     B&-&-&-&-&-&-&-&-&4.1e-19&1.1e-18\\
z=0.5  &A&-&-&-&-&-&-&-&3.2e-19& 5.7e-19 &1.5e-18\\
        &     B&-&-&-&-&-&-&-&1.4e-19 & 6.7e-19  &2.5e-19\\
z=1  &A&-&-&-&-&-&3.1e-19&5.1e-19&1.3e-19&2.2e-19&6.0e-19\\
        &     B&-&-&-&-&-&7.7e-20&1.3e-19&3.2e-20&5.5e-20&1.5e-19\\
z=2 & A &- &-&3.5e-20&8.9e-20&1.1e-19&1.5e-19&2.4e-19&6.1e-20&1.1e-19&2.9e-19\\
&  B &- &-&3.9e-21&1.0e-20&1.2e-20&1.7e-20&2.7e-20&6.8e-21&1.2e-20&3.2e-20\\
z=3 & A&1.4e-20&1.2e-19&2.5e-20&6.3e-20&7.6e-20&1.0e-19&1.7e-19&4.4e-20 &7.6e-20&2.1e-19\\
         &  B &8.7e-22&7.5e-21&1.6e-21&3.9e-21&4.7e-21&6.2e-21&1.1e-20&2.7e-21&4.8e-21&1.3e-20 \\
z=4  & A&1.2e-20&9.9e-20&2.1e-20&5.2e-20&6.3e-20&8.7e-20&1.4e-19&3.6e-20&6.3e-20&1.7e-19\\
       &B&4.8e-22&4.0e-21&8.4e-22&2.1e-21&2.5e-21&3.5e-21&5.6e-21&1.4e-21&2.5e-21&6.8e-21\\
z=5 & A&1.0e-20&8.5e-20&1.7e-20&4.5e-20&5.4e-20&7.5e-20&1.2e-19&3.1e-20&5.4e-20&1.4e-19\\
        & B&2.8e-22&2.4e-21&4.7e-22&1.2e-21&1.5e-21&2.1e-21&3.3e-21&8.6e-22&1.5e-21&3.9e-21\\
\hline
\hline
&&&&&&&&&&&\\
     &  &[SIII]&[SiII]&[NeIII]&[OIII]&[NIII]&[OI]&[OIII]&[NII]&[OI]&[CII]\\
$\lambda$($\mu$m)& & 33.5&	34.8&36.&52.&	57.&	63.&	88.&	122.&145.&157.\\
&Obs.&5.5e-15&9.1e-15&1.8e-15&1.1e-14&5.1e-15&1.6e-14&1.1e-14&3.0e-15&1.2e-15&2.2e-14\\
\hline
z=0.1  &A&6.9e-18&1.1e-17&2.3e-18&1.4e-17&6.4e-18&2.0e-17&1.4e-17&3.8e-18&1.5e-18&2.8e-17\\
            &B&5.7e-18&9.1e-18&1.9e-18&1.2e-17&5.3e-18&1.6e-17&1.2e-17&3.1e-18&1.2e-18&2.3e-17\\
z=0.2  &A&2.0e-18&3.3e-18&6.5e-19&4.0e-18&1.8e-18&5.8e-18&4.0e-18&1.1e-18&4.3e-19&7.9e-18\\
            &B&1.4e-18&2.3e-18&4.5e-19&2.8e-18&1.2e-18&4.0e-18&2.8e-18&7.6e-19&3.0e-19&5.5e-18\\
z=0.3  &A&1.0e-18&1.6e-18&3.2e-19&2.0e-18&9.2e-19&2.9e-18&2.0e-18&5.4e-19&2.2e-19&4.0e-18\\
            &B&5.9e-19&9.5e-19&1.9e-19&1.2e-18&5.4e-19&1.7e-18&1.2e-18&3.2e-19&1.3e-19&2.4e-18\\
z=0.4  &A&6.3e-19&1.0e-18&2.0e-19&1.2e-18&5.8e-19&1.8e-18&1.2e-18&3.4e-19&1.4e-19&-\\
            &B&3.2e-19&5.1e-19&1.0e-19&6.0e-19&2.9e-19&9.2e-19&6.1e-19&1.7e-19&7.1e-19&-\\
z=0.5  &A&4.4e-19&7.3e-19&1.4e-19&8.9e-19&4.1e-19&1.3e-18&8.9e-19&2.4e-19&-&-\\
            &B&1.9e-19&3.2e-19&6.2e-20&4.0e-19&1.8e-19&5.8e-19&4.0e-19&1.1e-19&-&-\\
z=1     &A&1.7e-19&2.9e-19&5.6e-20&3.5e-19&1.6e-19&5.1e-19&3.5e-19&-&-&-\\
            &B&4.2e-20&7.2e-20&1.4e-20&8.7e-20&4.0e-20&1.3e-19&8.7e-20&-&-&-\\
z=2    &A&8.4e-20&1.4e-19&2.7e-20&1.7e-19&7.7e-20&2.4e-19&-&-&-&-\\
           &B&9.3e-21&1.6e-20&3.0e-21&1.9e-20&8.6e-21&2.7e-20&-&-&-&-\\
z=3    &A&6.0e-20&9.9e-20&1.9e-20&1.2e-19&-&-&-&-&-&-\\
           &B&3.8e-21&6.2e-21&1.2e-21&7.5e-21&-&-&-&-&-&-\\	
z=4    &A&4.9e-20&8.1e-20&1.6e-20&-&-&-&-&-&-&-\\
           &B&2.0e-21&3.2e-21&6.4e-22&-&-&-&-&-&-&-\\
z=5    &A&4.2e-20&7.0e-20&1.4e-20&-&-&-&-&-&-&-\\
           &B&1.2e-21&1.9e-21&3.9e-22&-&-&-&-&-&-&-\\	
\hline
\hline



\end{tabular}
\end{center}
{ {\bf Notes}: for each transition in each column, the first line gives the observed intensity of NGC1068, the other lines list for the given redshift the predicted intensity value for models A (top value) and B (bottom value). The blank values correspond to observed wavelengths outside the ESI-SPICA range $\sim$30-200$\mu$m.}
\end{table*}

\section{Detection of the first stars}

Primordial stars are born in a Universe which was very poor in metals, for this reason they should form in very massive clouds which are expected to originate in very massive clusters populated by a large number of massive stars. Following the suggestion by Panagia (2005) to detect the primordial stars though their HII regions nebular emission, we modeled the emission of primordial clusters using the model stellar atmosphere by Schaerer \& de Koter (1997) and a standard photoionization model.

We have used a stellar model with mass M=120 M$_{\odot}$ (the largest value of the mass among the models by Schaerer \& de Koter), effective temperature
T$_{eff}$= 53,400K, luminosity  L=2 $\times$ 10$^{6}$ L$_{\odot}$, and radius R=15.6 R$_{\odot}$. 
We computed the spectrum of an HII region originated from this star, using the CLOUDY v96 photoionization code with the following parameters: internal radius R(int)=0.1pc, external radius R(ext)=2.5pc, metallicity of 10$^{-6}$ Z$_{\odot}$, Hydrogen density of 10$^{3}$ cm$^{-3}$.

We found that to obtain flux densities comparable to the spectroscopic detection limit of JWST, about 10$^5$ stars are needed in the cluster.
We present in Fig.3 the synthetic spectrum of such a cluster at a redshift of z=15. It can be seen from the figure that such clusters, if they existed in the primordial Universe, can be detected through 
Hydrogen and Helium recombination lines by JWST in the mid-IR (e.g. in the Ly$\alpha$, H$\alpha$, He I 1.08$\mu$m lines) and by SPICA in the far-IR (e.g. in the Pa$\alpha$ and  Br$\alpha$ lines). 
The lines that can be observed in the far-IR are better than the Ly$\alpha$ line, because this latter can be severely absorbed by the dust, which could be present in the HII region, if one allows moderate metallicities of the order of  10$^{-3}$ Z$_{\odot}$ (Panagia 2005). Moreover, the Ly$\alpha$ line can also be absorbed by HI in the intergalactic medium (Miralda-Escude \& Rees 1998).

The sensitivity of a mid-to-far-IR spectrometer onboard of SPICA should be better than that one of JWST, at wavelenths larger than 10 $\mu$m, because it will not be limited by the thermal background of the mirror.
We conclude that the estimated number of members of the cluster can well decrease by a factor 10, to the value of 10$^4$ stars, if we consider stars with a larger mass, e.g. of the order of 150-200 M$_{\odot}$, for which reliable models are not available. 

\begin{figure}[ht]
\centerline{\includegraphics[angle=0,width=13cm]{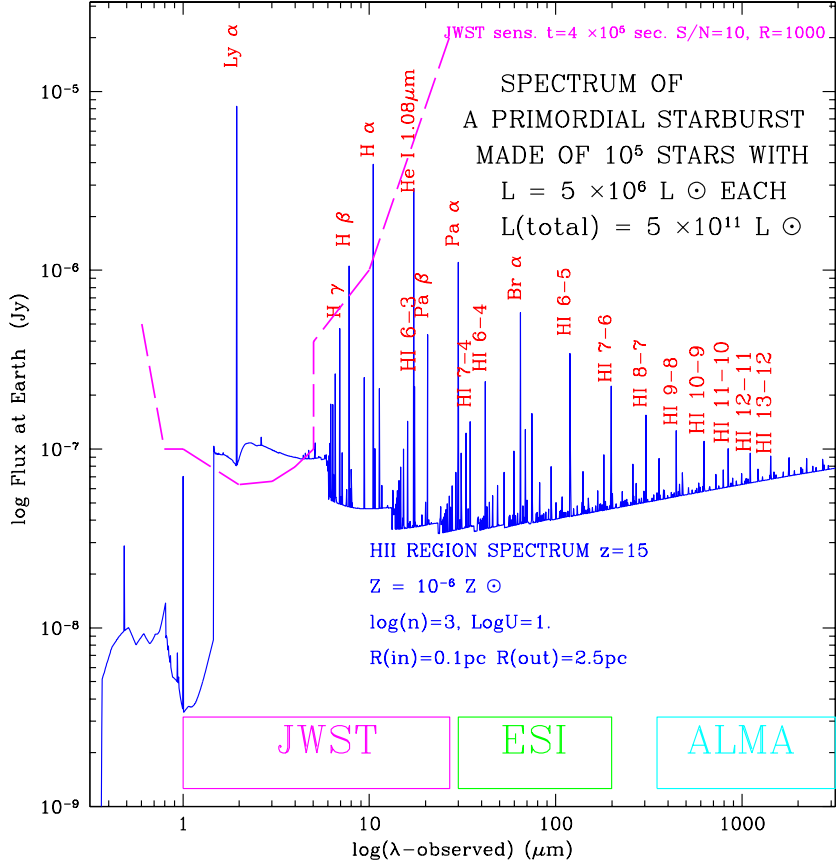}}
\caption{ \footnotesize Redshifted (z=15) starburst spectrum of 10$^5$ stars ionizing HII regions
 Z = 10$^{-6}$ Zo. Flux is scaled at Earth. The spectral ranges of JWST, ESI onboard of SPICA and ALMA are indicated. The spectroscopic sensitivity of JWST is given for t=4$\times$10$^5$ sec., S/N=10, R=1000.} \label{fig3}
\end{figure}

\end{document}